\documentclass[aps,pre,reprint,groupedaddress]{revtex4-1}
\usepackage{graphicx}
\usepackage{dcolumn}
\usepackage{bm}
\usepackage{epsfig}
\usepackage{subfigure}
\usepackage{graphics}
\usepackage{epstopdf}
\usepackage[english]{babel}
\usepackage{amsfonts}
\usepackage{mathrsfs}
\usepackage{amsthm}
\usepackage{amssymb}
\usepackage{enumerate}
\usepackage{amsmath}
\usepackage{subfigure}

\begin{document}


\title{Exploring the Node Importance Based on von Neumann Entropy}


\author{Xiangnan Feng$^{1,2}$}
\author{Wei Wei$^{1,2,3}$}
\email[]{weiw@buaa.edu.cn}
\author{Jiannan Wang$^{1,2}$}
\author{Ying Shi$^{1,2}$}
\author{Zhiming Zheng$^{1,2,3}$}

\affiliation{$^{1}$School of Mathematics and Systems Science, Beihang University, Beijing, China\\
$^{2}$Key Laboratory of Mathematics Informatics Behavioral Semantics, Ministry of Education, China\\
$^{3}$Beijing Advanced Innovation Center for Big Data and Brain Computing, Beihang University, Beijing, China}


\date{\today}

\begin{abstract}
When analyzing the statistical and topological characteristics of complex networks, an effective and convenient way is to compute the centralities for recognizing influential and significant nodes or structures, yet most of them are restricted to local environment or some specific configurations. In this paper we propose a new centrality for nodes based on the von Neumann entropy, which allows us to investigate the importance of nodes in the view of spectrum eigenvalues distribution. By presenting the performances of this centrality with network examples in reality, it is shown that the von Neumann entropy node centrality is an excellent index for selecting crucial nodes as well as classical ones. Then to lower down the computational complexity, an approximation calculation to this centrality is given which only depends on its first and second neighbors. Furthermore, in the optimal spreader problem and reducing average clustering coefficients, this entropy centrality presents excellent efficiency and unveil topological structure features of networks accurately. The entropy centrality could reduce the scales of giant connected components fastly in Erd\"os-R\'enyi and scale-free networks, and break down the cluster structures efficiently in random geometric graphs. This new methodology reveals the node importance in the perspective of spectrum, which provides a new insight into networks research and performs great potentials to discover essential structural features in networks.
\end{abstract}


\maketitle

\section{Introduction}
Networks provide us a useful tool to analyze a wide range of complex systems, including WWW \cite{6621057}, the social structure \cite{wasserman1994social}, the economic behaviors \cite{cano2006topology}, and the biochemical reactions \cite{zitnik2018modeling}. Since the 1990s, a great number of interdisciplinary studies involving network both in theories and empirical work, have come up and developed new models and techniques to shed a light on the complex structure behind the particular subjects.

Among these studies, centralities which indict the most important nodes in networks have received considerable attention. Many centralities have been proposed to describe and measure the importance of nodes in certain aspect to perform their specific explanations for importance. The closeness centrality \cite{bavelas1950communication,Sabidussi1966} of a node is defined as the average shortest pathes from this node to all the others in the whole network and aims to find the center based on pathes and node geological positions. Freeman \cite{freeman1977set} introduced the betweenness centrality to measure the controlling ability of nodes on communication between each pair of nodes and help find the nodes that control the information flow in the network. Bonacich \cite{10.2307/2780000} proposed the eigenvector centrality which takes into account the influence of powerful neighbors when evaluating node importance, and based on this the PageRank is established by Larry Page et al.\ \cite{page1999pagerank} to work as the core websites-ranking algorithm of Google. Piraveenan et al.\ \cite{piraveenan2013percolation} proposed the percolation centrality to measure the node importance in aiding percolation of networks and considered that the values of nodes centrality depend on their states, which is widely implemented in percolation networks like contagion and computer virus spreading processes. These methods promote the application of network researches and deeper our understanding in complex systems.

However, when measuring node importance, one single centrality is not always perfect since all the centralities only describe networks by a specific perspective. A huge number of real-world data is complex and requires multiple or comprehensive description. For example in a social network \cite{leskovec2012learning}, nodes with high degrees or high betweenness centrality could have significant yet different effects on the network. In this situation it is necessary to make sure that an all-round description about the nodes importance could be presented.

Till now, most studies on complex network focus on graph theory, which mostly focuses on local structure and heuristic strategies such as centrality and modularity, and entropy provides an alternative way to measure the global characterization and had won great success in many researching fields. The von Neumann entropy (or quantum entropy) has shown great success in qualifying the organization structure and levels in networks, and can be applied in networks as an index to quantify the network heterogeneous characteristics. Passerini et al.\ \cite{2008arXiv0812.2597P} used the normalized combinational Laplacian matrix of networks to study the quantum state and von Neumann entropy of networks, and proved that the regular graphs and complete graphs have maximum entropy while networks with the same number of nodes and edges which contain large cliques have the minimum entropy. According to this result the von Neumann entropy could reflect the regularity of networks. By defining a rank-1 operator in the bipartite tensor product space \cite{2013arXiv1304.7946D}, Beaudrap et al.\ provided an interpretation of von Neumann entropy, and regarded the von Newmann entropy as the measurement of quantum entanglement between two systems corresponding to edges and nodes respectively. Han et al.\ \cite{Han20121958} developed a simplified von Neumann entropy which could be computed using nodes degree statistics, compared it with Estrada's heterogeneity index of node \cite{estrada2010quantifying}, and concluded that the von Neumann entropy can be used to measure the network complexity. The von Neumann entropy describes networks integrally and allows us to combine it with other concepts.

To analyze data and understand the inherent structure and organization of networks better, we make use of the von Neumann entropy to establish a new measurement for centrality. In section $2$ the von Neumann entropy of networks is introduced, and the corresponding node centrality for networks is defined. Then some examples are analyzed and the specific calculation related to the entropy, including approximation, is presented. Next in section $3$ some experiments are implemented and the von Neumann entropy centrality is compared with other centralities. The behaviors of the entropy centrality in optimal influencers problem, average clustering coefficient and spearman correlation coefficient are shown. These researches build the centrality with spectrum of networks and demonstrate their superior in describing the importance of nodes in a new perspective.
\section{Von Neumann Entropy and Centrality}
\subsection{Spectral Distribution of Eigenvalues}
Given an undirected network $G(V,E)$, $V$ (or $V(G)$) is a finite set whose elements are nodes of the network $G$ and $E$ (or $E(G)$) is the edges set. $E$ is composed of unordered pairs of nodes who belong to $V$, namely, when $(v_i,v_j)\in E$, we have $(v_j,v_i)\in E$ and $v_i,v_j\in V$. The edge in the form of $(v_i,v_i)$ is called a self-loop. In this paper we only talk about the networks without self-loops. The \emph{adjacency matrix} is an $N\times N$ matrix, where $N = |V|$. Using $A(G)$ to denote the adjacency matrix of $G$, the columns and rows of $A(G)$ are labeled by the vertices of $G$, and the $(i,j)$ entry of $A(G)$ is $1$ if and only if $(v_i,v_j)\in E(G)$, namely the adjacency matrix $A(G)$ could be defined as follows:
\begin{eqnarray}
[A(G)]_{i,j}=\left\{
\begin{array}{ll}
1 & \textrm{if $(v_i,v_j)\in E$,}\\
0 & \textrm{if $(v_i,v_j)\notin E$}.
\end{array} \right.
\end{eqnarray}

Before the introduction of von Neumann entropy, firstly the normalized Laplacian matrix is introducted \cite{chung1997spectral}. The degree of a vertex $v_i\in G$, denoted as $d_G(v_i)$ or $d_i$, is the total number of edges touching this vertex. In this way we could define the \emph{degree matrix} which is an $N\times N$ diagonal matrix and denoted as $D(G)$. The entries in the degree matrix are defined as follows:
\begin{eqnarray}
[D(G)]_{i,j}=\left\{
\begin{array}{ll}
d_G(v_i) & \textrm{if $i=j$,}\\
0 & \textrm{if $i\neq j$}.
\end{array} \right.
\end{eqnarray}
The \emph{combinatorial Laplacian matrix} $L(G)$ could be define as $L(G)=D(G)-A(G)$:
\begin{eqnarray}
[L(G)]_{i,j}=\left\{
\begin{array}{ll}
d_G(v_i) & \textrm{if $i=j$,}\\
-1 & \textrm{if $i\neq j$, $(v_i,v_j)\in E$,}\\
0 & \textrm{otherwise}.
\end{array} \right.
\end{eqnarray}

It is worth noting that the Laplacian matrix will not change if the self-loop is added or deleted. As we can see, the Laplacian matrix is a diagonally dominant Hermite matrix, thus it is positive semi-defined \cite{Horn:2012:MA:2422911}.

The \emph{normalized Laplacian matrix} is defined as $\mathcal{L} = D^{-1/2}LD^{-1/2}$ and the elements are:
\begin{eqnarray}
[\mathcal{L}(G)]_{i,j}=\left\{
\begin{array}{ll}
1 & \textrm{if $i=j$, $d_G(v_i)\ne 0$,}\\
-\frac{1}{\sqrt{d_G(v_i)d_G(v_j)}} & \textrm{if $i\neq j$, $(v_i,v_j)\in E$,}\\
0 & \textrm{otherwise}.
\label{eq:one}
\end{array} \right.
\end{eqnarray}
The spectral decomposition of $\mathcal{L}(G)$ is $\mathcal{L}(G) = \Phi\Lambda\Phi$, where $\Lambda = diag(\lambda_1,\lambda_2,\dots,\lambda_N)$ is a diagonal matrix of eigenvalues with order $0=\lambda_1\leq\lambda_2\leq \cdots \leq\lambda_N$ and $\Phi$ is a matrix whose columns are orthonormal eigenvectors corresponding to the ordered eigenvalues. Notice that the normalized Laplacian matrix is also semi-defined so all the eigenvalues are non-negative.

When it comes to measure the importance of nodes in a network, many classical properties, like degree and clustering coefficient, are local parameters and only capture microscopic features. Some other indices like betweenness and closeness centralities only depict specific parts of the behavior of nodes and do not perform integral properties of networks. Global and computational efficient parameters are needed to help measure the complexity/importance and unveil the difference of nodes. Since there exists an isomorphism between normalized Laplacian matrices and networks, it could be concluded that all the information of a network is contained in its Laplacian matrix and the Laplacian matrix holds the complete characterizations of the network. Researchers have found a lot of indices and parameters to measure the structural complexity and regularity of networks based on the Laplacian matrix, which could be viewed as a data reduction from $N\times N$ matrix to an index. Xiao et al.\ \cite{Xiao20092589} explored to use the trace of heat kernel to measure similarity and clustering of networks. Estrada \cite{estrada2010quantifying} designed a heterogeneity index based on the variation of degree functions of all pairs of linked nodes, and gave bounds for this index when quantifying the heterogeneity of different networks.

For the Laplacian matrix, the most important indices are the eigenvalues and they are directly related to the topological properties of the network: the number of eigenvalues equalling to zero is the number of connected components in this network and there is only one zero-eigenvalue for a connected network; $\lambda_2$ is referred as the algebraic connectivity and the corresponding eigenvector is known as Fiedler vector \cite{fiedler1973algebraic} \cite{fiedler1989laplacian}, which is frequently used to network partition \cite{powers1988graph}. By the way, for the normalized Laplacian matrix, all the eigenvalues satisfy $0\leq\lambda_i\leq2$, $1\leq i\leq N$ and the upper limit $2$ is achieved only when the network is bipartite. Evaluating the accurate ranges of each eigenvalue is still an open problem.

\begin{figure}[htbp]
\begin{center}
\includegraphics[scale=0.18]{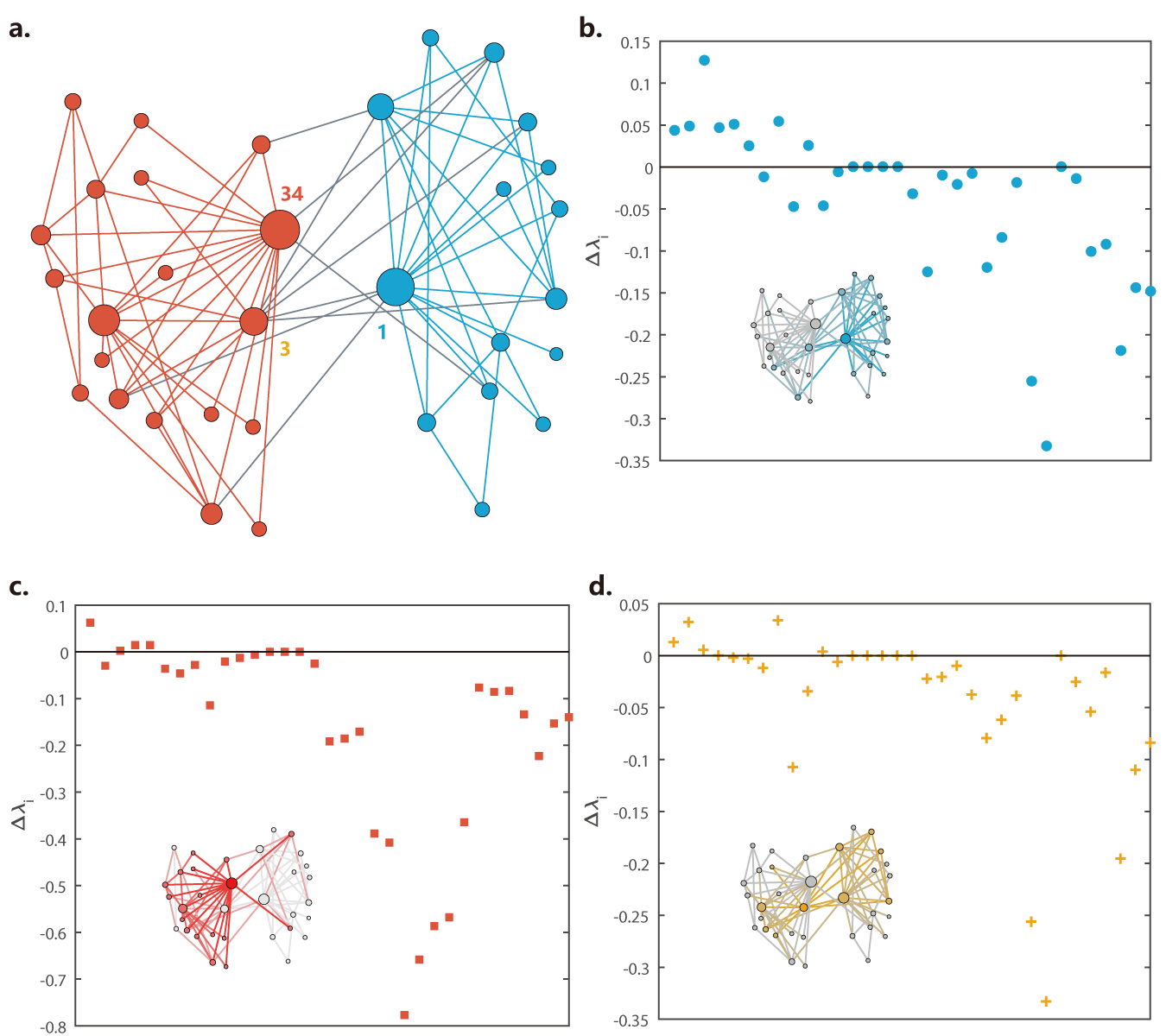}
\caption{\label{Fig:1} Zachary's karate club network. There are 34 members in the club and 78 links outside the club. Node 1 is the instructor and node 34 is the club administrator or president. A conflict has happened between the instructor and administrator during the study, which led to the split of the club. We could see that removing different nodes will bring different changes to the spectrum. \textbf{(a.)} The connection relationship between the numbers. \textbf{(b.)} The changes of eigenvalues after node 1 is removed. The eigenvalues are sort in decreasing order and the points in the graph stands for the variations of eigenvalues. \textbf{(c.)} The changes of eigenvalues after node 34 is removed. \textbf{(d.)} The changes of eigenvalues after node 3 is removed.}
\end{center}
\end{figure}

Since the eigenvalues are the most significant features of a matrix, we believe the distribution of the spectrum (distribution) is fatal to a network concerning the topological characteristics. The normalized Laplacian matrix contains all the topological characteristics of a network, thus the spectrum eigenvalues could be viewed as a natural data reduction of information in statistical viewpoint. If different nodes are removed from the network, various changes will be brought to its spectrum distribution. An example of Zachary's karate club network \cite{zachary1977information} is show in Figure \ref{Fig:1}. Considering the roles played by node 34 and node 1 in the club, they are more significant than node 3, thus removing node 34 or 1 will bring larger changes on the spectrum than removing node 3. Capturing the variations in spectrum distribution will lead to a significant understanding in structural changes of the network.

\subsection{Von Neumann Entropy Centrality}
As a crucial way of depicting distributions, entropy could be used to signify the features of spectrum distribution \cite{chung1997spectral}. The von Neumann entropy, commencing from normalized Laplacian matrix, could be regarded as a well-designed and sophisticated representation of network. There are many researches related to von Neumann entropy and its function in describing the network structure, which receive quite a lot of attention in many applications. This index integrates the complete values and properties of all the eigenvalues and thus could reflect global structural complexity and characteristics.

The von Neumann entropy of a network $G$ associated with its normalized Laplacian matrix $\mathcal{L}(G)$, denoted as $S(G)$, is defined as \cite{2008arXiv0812.2597P} \cite{chung1997spectral} \cite{han2011characterizing}:
\begin{eqnarray}
S(G)=-\sum_{i=1}^{N}\frac{\lambda_i}{2}\ln{\frac{\lambda_i}{2}},
\end{eqnarray}
where $\lambda \log\lambda=0$ when $\lambda =0$.

Accordingly, the centrality of node $v$ can be defined as the variation of von Neumann Entropy when removing this node and edges linked to it from the network. Using $C_{E}(v)$ to denote the \emph{von Neumann entropy node centrality} of node $v$, we have
\begin{eqnarray}
C_{E}(v)=|S(G)-S(G\setminus v)|.
\end{eqnarray}
Similarly, this centrality could be defined on other structures in graph. Let $s$ be a subnetwork of $G$, and denote $G\setminus s$ to be the network remained after deleting the nodes in $s$ and edges linked with these nodes. The \emph{von Neumann entropy centrality} of subnetwork $s$ could be defined as:
\begin{eqnarray}
C_{E}(s)=|S(G)-S(G\setminus s)|.
\end{eqnarray}

According to known researches related to von Neumann entropy \cite{Han20121958}, $S(G)$ can reflect the regularity and complexity of the network and is effective to characterize the global structure of networks. When a node is removed, the network will change, which leads to the change of the Laplacian matrix and its eigenvalues, thus the von Neumann entropy of the network will finally change. If deleting node $x$ brings larger change of $S(G)$ than deleting node $y$, it proves that deleting node $x$ could cause more significant change on the spectrum distribution and network structure. Since the relationship between the network and its spectrum is elaborate and profound, when removing a node from the network, the change in von Neumann entropy will accurately present the impact of this node on the whole network structure, which makes the von Neumann entropy a great parameter to build centrality.

\subsection{Some Examples}
\begin{figure}[htbp]
\begin{center}
\includegraphics[scale=0.39]{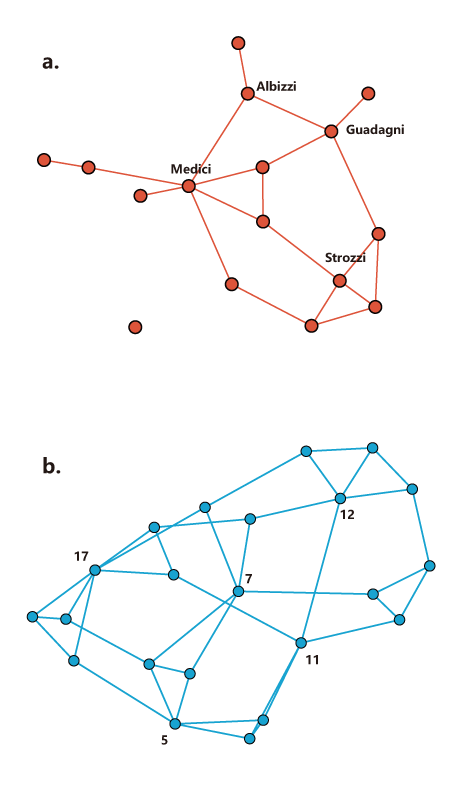}
\caption{\label{Fig:2} \textbf{(a.)} Padgett Florentine families marital ties network. $16$ nodes and $20$ edges are contained in this network. The Pucci family did not have martial tie with others, so the major part of the network is a component with $15$ nodes. \textbf{(b.)} The gift-exchange network in a Papuan village. There are $22$ nodes and $39$ edges. Each node stands for a household and each edge stands for gift exchange.}
\end{center}
\end{figure}

To illustrate the efficiency of von Neumann entropy centrality, some specific networks are used to perform and compare different node centralities in this subsection. The results of betweenness centrality (BC), closeness centrality (CC), degree centrality (DC) and von Neumann entropy node centralities on Padgett Florentine families network and the gift-giving network are shown in Figure \ref{Fig:2}.

The Padgett Florentine families network is a network of marital ties among Renaissance Florentine families \cite{breiger1986cumulated, doi:10.1177/000271628145500136}. This network is built based on historical documents and an edge between two nodes means there existed marriage alliance between the two corresponding families. The network includes families who were involved in the struggle for the control of the city in politics around 1430s. $16$ families are contained in the network and there is a major component consisted by $15$ of them. The ranks of each centralities is shown in Table \ref{tab:1}. As we could see, all the node centralities rank the Medici as the most influential one. This actually coincides with historical fact since the Medici family is one of the most famous families in history who reached peak in Italian upper classes during Renaissance. The von Neumann entropy centrality is able to find out the most influential families correctly as other node centralities: all the first three families in the $C_{E}$ sort appear in other three centrality sorts and the Medici family has the highest centrality. This reflects that the von Neumann entropy centrality could work as a new reasonable and accurate centrality and is able to exactly capture the nodes which work as the most influential ones and are crucial to the whole network.

\begin{table}[htbp]
\caption{\label{tab:1}
Ranks of nodes of Padgett Florentine families marital ties network in BC, CC, DC and $C_E$.}
\begin{ruledtabular}
\begin{tabular}{@{}*{5}{c}}
\textrm{Ranks}&
\textrm{BC}&
\textrm{CC}&
\textrm{DC}&
\textrm{$C_E$}\\
\colrule
1st & Medici & Medici & Medici & Medici\\
2nd & Guadagni & Ridolfi & Guadagni & Guadagni\\
 & & & Strozzi & \\
3rd & Albizzi & Albizzi & Albizzi & Albizzi\\
 & &Tornabuon & Bischeri ... &\\
\end{tabular}
\end{ruledtabular}
\end{table}
The other example, the gift-giving network, shows the gift exchange relations among $22$ households in a Papuan village \cite{nla.cat-vn2294502, Schwimmer_1970}. In this network, if two households exchange gifts, there will be an edge between them. In this village, the gift-exchanging is significant in life because it is regarded as a method to request political and economic assistance from others and works as the pristine market. Although there may exist deep contents and meanings behind the whole process in the network, yet it is natural to realize that the family who exchanges gifts with more persons and have higher degrees may have larger influence on the whole village. At the same time, since the exchange process could be long and complicated, like the family $A$  may ask family $B$ to ask family $C$ to assist $A$, the betweenness centrality and closeness centrality will also point out influential households or persons in the network. Thus it is incomplete to evaluate the network with only one single centrality. Multiple centralities are required to help understand the structure and information behind the network better. As shown in the Table \ref{tab:2}, the entropy centrality performs its potential to work as an all-round index on node-importance: the first five nodes in $C_{E}$ sort have highest ranks in other centralities sorts, like node $11$ ranks first in BC sort and CC sort, node $12$ ranks the second in DC sort. The von Neumann entropy centrality could be viewed as a combination of other centralities and the all-round property of $C_{E}$ allows us to find more meaningful information in the network.
\begin{table}[htbp]
\caption{\label{tab:2}
Ranks of nodes of gift-giving network in BC, CC, DC and $C_E$.
}
\begin{ruledtabular}
\begin{tabular}{@{}*{5}{c}}
\textrm{Ranks}&
\textrm{BC}&
\textrm{CC}&
\textrm{DC}&
\textrm{$C_E$}\\
\colrule
1st & 11 & 11 & 17 & 17\\
2nd & 7 & 7 & 5/7/11/12 & 11\\
3rd & 17 & 17/19 & 4 & 12\\
4th & 12 & 12/16 & 1/2/3... & 7\\
5th & 5 & 4/18 &  & 5\\
\end{tabular}
\end{ruledtabular}
\end{table}

There are more interesting things in the gift-giving network. A natural question is since node 7 ranks higher than 12 in both BC and CC and they rank the same in DC, why node 12 ranks higher than node $7$?
\begin{figure}[htbp]
\begin{center}
\includegraphics[scale=0.28]{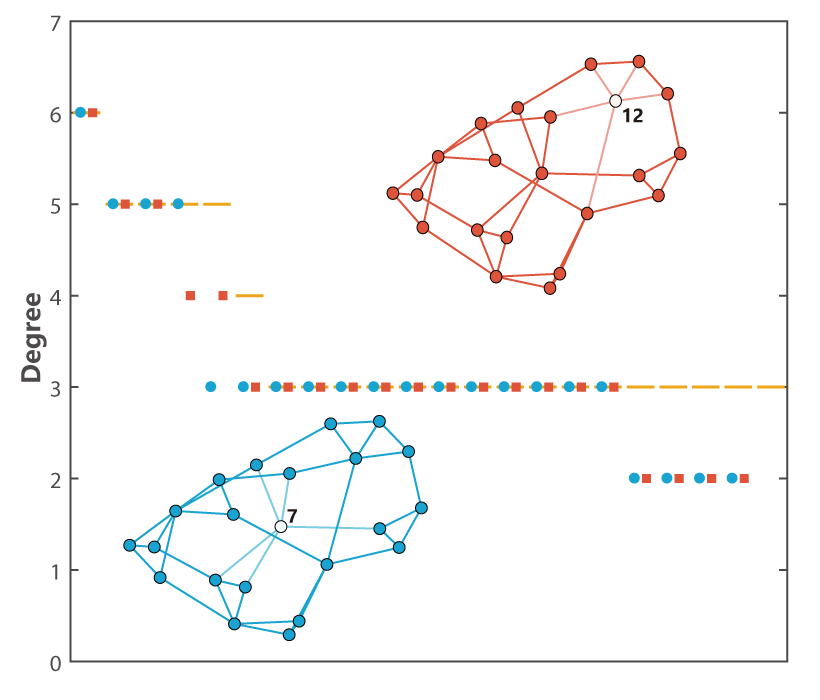}
\caption{\label{Fig:3} The degree distributions after the removal of node 7 or node 12. The oranges lines are the degrees of original network. Blue circles or red squares are the degrees when node 7 or node 12 is removed. The degrees in each network are sorted in decreasing order.}
\end{center}
\end{figure}

We infer that this may result from the topological changes in the graph when the nodes are removed. Node 12 is linked with the hub 11 and when node 12 or node 7 is removed, the degrees of the remained graphs are shown in table \ref{tab:3} and Figure \ref{Fig:3}. Using $-\sum_{i}p_i\log{p_i}$ to calculate the entropy of the degree distributions, the entropy of the original graph is $0.8226$. While node $12$ or $7$ is removed, the entropy is changed into $1.2285$ and $1.0357$. Since the entropy indicates the irregularity of corresponding distribution $p_i$, it is suggested that the removal of node $12$ brings larger variations in the degree distributions and makes the network more even. This helps explain why node 12 ranks higher than node 7.
\begin{table}[htbp]
\caption{\label{tab:3}
Degree distributions when node 7 or node 12 is removed from the gift-giving network.
}
\begin{ruledtabular}
\begin{tabular}{@{}*{6}{c}}
\textrm{Degrees}&
\textrm{6}&
\textrm{5}&
\textrm{4}&
\textrm{3}&
\textrm{2}\\
\colrule
Original & 1 & 4 & 1 & 16 & 0\\
Remove Node 12 & 1 & 2 & 2 & 12 & 4\\
Remove Node 7 & 1 & 3 & 0 & 13 & 4\\
\end{tabular}
\end{ruledtabular}
\end{table}

From networks above, the von Neumann entropy centrality is a combination of traditional centralities and could be viewed as a comprehensive measure of node importance. The $C_{E}$ takes the global network structure into account and has excellent performance in selecting significant nodes in network.

\subsection{Approximation to Entropy}
To calculate all or most of the eigenvalues of the matrix $\mathcal{L}$, a number of algorithms are studied. By a similarity transformation with orthogonal matrix $Q$, the matrix $\mathcal{L}$ could be transformed to an upper triangular matrix $T = Q^T\mathcal{L}Q$ where $T$ and $\mathcal{L}$ own the same eignevalues. Eigenvalues of $T$ could be calculated by methods like QL algorithm \cite{Press:1992:NRC:148286} with complexity $O(n)$. The orthogonal matrix $Q$ could be calculated by various methods. For a symmetric matrix, the $Q$ could be researched by Householder algorithm \cite{Householder:1958:UTN:320941.320947} with complexity $O(n^3)$. For a sparse matrix, Lanczos algorithm \cite{lanczos1950iteration} could find $Q$ with complexity $O(mn)$ where $m$ is the edge number of the network.

However, since in reality the scale of networks could be enormous, the complete algorithms above are not applicable considering the time consuming. To efficiently apply the entropy centrality, the approximation of entropy will be discussed in this subsection.

Expanding at $x=1$, we could easily get that
\begin{eqnarray}
\ln(x)=x-1-\sum_{k=2}^{\infty}\frac{(1-x)^k}{k}.
\label{eq:two}
\end{eqnarray}
This series could be applied to approximate the entropy by cutting off at some index $k$, and we use $\ln(x) = x-1$ to approach the entropy. In this situation, the entropy is calculated as:
\begin{eqnarray}
S &&=-\sum_{i=1}^{N}\frac{\lambda_i}{2}\ln{\frac{\lambda_i}{2}}\backsimeq \sum_{i=1}^{N}\frac{\lambda_i}{2}(\frac{\lambda_i}{2}-1)\nonumber\\
&& =\frac{1}{2}\sum_{i=1}^{N}\lambda_i-\frac{1}{4}\sum_{i=1}^{N}\lambda_i^2.
\end{eqnarray}
Since $Tr(\mathcal{L}^n)=\sum_{i}(\lambda^n_i)$, the approximated entropy could be written as:
\begin{eqnarray}
S_1 = \frac{1}{2}Tr(\mathcal{L})-\frac{1}{4}Tr(\mathcal{L}^2).
\end{eqnarray}
According to the definition in equation (\ref{eq:one}), $Tr(\mathcal{L}) = |V|$.

To calculate $Tr(\mathcal{L}^2)$, with some linear algebra knowledge, it is concluded that
\begin{eqnarray}
Tr(\mathcal{L}^2) &&= Tr(\mathcal{L}\times\mathcal{L}) = \sum_{i}\sum_{j}\mathcal{L}_{ij}\mathcal{L}_{j,i}\nonumber\\
&& = \sum_{i}\sum_{j}\mathcal{L}^2_{ij} = \sum_{i=j}\mathcal{L}^2_{ij} + \sum_{i\neq j}\mathcal{L}^2_{ij}\nonumber\\
&& = |V| + \sum_{i\sim j}\frac{1}{d_id_j}.
\end{eqnarray}
In this way, the von Neumann entropy is approximated as:
\begin{eqnarray}
S_1 &&= \frac{|V|}{2} - \frac{|V|}{4} -\sum_{i\sim j}\frac{1}{4d_id_j}\nonumber\\
&& = \frac{|V|}{4} - \sum_{i\sim j}\frac{1}{4d_id_j}.
\end{eqnarray}
By this calculation, let the network with node $v_i$ removed be $G'$, the von Neumann entropy centrality of node $v_i$ is
\begin{eqnarray}
C_E(v_i) && \backsimeq |S_1(G)-S_1(G')|\nonumber\\
&& = |\frac{|V(G)|}{4} - |\frac{|V(G')|}{4}| - \sum_{j\sim k}\frac{1}{4d_jd_k} + \sum_{j\sim k}\frac{1}{4d'_jd'_k}\nonumber\\
&& = |\frac{|V(G)|-|V(G')|}{4}| - \sum_{j,i\sim j}\frac{1}{4d_id_j}\nonumber\\
&& \quad -\sum_{j,k,i\sim j\sim k}\frac{1}{4d_jd_k} + \sum_{j,k,i\sim j\sim k}\frac{1}{4d'_jd'_k}
\end{eqnarray}
If node $v_j$ is linked with $v_i$, then $d'_j=d_j-1$. Hence,
\begin{eqnarray}
C_E(v_i) &&\backsimeq \frac{1}{4}- \sum_{j,i\sim j}\frac{1}{4d_id_j}\nonumber\\
&& + \sum_{j,k,i\sim j\sim k}\frac{1}{4(d_j-1)d_jd_k}
\end{eqnarray}

To cut off the series in equation (\ref{eq:two}) at a higher $k$ could help improve the accuracy. In order to calculated $\sum_{i}\lambda_i(1-\lambda_i)^k$, the sum of eigenvalues with higher power $\sum_{i}\lambda^t_i = Tr(\mathcal{L}^t)$ for $2\leq t\leq k+1$ need to be solved. They could be calculated in other perspective. Taking $Tr(\mathcal{L}^3)$ as example, since
\begin{eqnarray}
\mathcal{L} = D^{-1/2}LD^{-1/2} = I - D^{-1/2}AD^{-1/2},
\end{eqnarray}
it could be got easily that
\begin{eqnarray}
&& \quad Tr[(I-\mathcal{L})^3]\nonumber\\
&& = Tr(D^{-1/2}AD^{-1/2}D^{-1/2}AD^{-1/2}D^{-1/2}AD^{-1/2})\nonumber\\
&& = Tr(D^{-1/2}AD^{-1}AD^{-1}AD^{-1/2})\nonumber\\
&& = \sum_i\sum_j\sum_k\frac{1}{\sqrt{d_i}}A_{ij}\frac{1}{d_j}A_{jk}\frac{1}{d_k}A_{ki}\frac{1}{\sqrt{d_i}}\nonumber\\
&& = \sum_i\sum_j\sum_k\frac{1}{d_id_jd_k}A_{ij}A_{jk}A_{ki}.
\end{eqnarray}
Since
\begin{eqnarray}
Tr[(I-\mathcal{L})^3] && = Tr(I-3\mathcal{L}+3\mathcal{L}^2-\mathcal{L}^3)\nonumber\\
&& = Tr(I-3\mathcal{L}) + 3Tr(\mathcal{L}^2) - Tr(\mathcal{L}^3),
\end{eqnarray}
then we have
\begin{eqnarray}
Tr(\mathcal{L}^3) && = -2|V| + 3|V| +\sum_{i\sim j}\frac{3}{d_id_j}\nonumber\\
&& \quad - \sum_i\sum_j\sum_k\frac{1}{d_id_jd_k}A_{ij}A_{jk}A_{ki}\nonumber\\
&& = |V| + \sum_{i\sim j}\frac{3}{d_id_j} -\sum_{i\sim j \sim k\sim i}\frac{1}{d_id_jd_k}.
\end{eqnarray}
So the approximated entropy when cutting off at $k=2$ in equation (\ref{eq:two}) is
\begin{eqnarray}
S_2(G)= \frac{5}{16}|V|-\sum_{i\sim j}\frac{11}{16d_id_j} + \sum_{i\sim j \sim k\sim i}\frac{1}{16d_id_jd_k}.
\end{eqnarray}
Similar derivation could be applied to the situation when $k>2$.

By a breadth-first search algorithm \cite{cormen2001introduction}, the neighbor-relation of nodes in a network could quickly be achieved. Then to calculate the entropy centrality of a node, we only need to consider all the paths starting from this node whose lengths are less than the order of cutting-off for approximation. If the neighbor-relations for each node are stored well, the calculation complexity will be reduced hugely and the global calculation is simplified and degenerated to local situations.

\section{Experiments}
To further explore the properties and features of von Neumann entropy centrality, we discuss the performance of this centrality in ``optimal influencer'' problem, namely finding the minimal set of nodes which are crucial in spreading information. Then the variations in average clustering coefficient will be presented. Finally the Spearman correlation coefficients between entropy centrality and others will be analyzed. Several centralities will be compared to perform and support our points.
\subsection{Minimal Set of Influencers}
Finding the minimal set of influencers in a network to spread information, or optimal percolation, is a widely researched problem\cite{granovetter1978threshold, watts2002simple}. It is believed that the most significant influencers are the nodes which are capable of preserving the connection of the whole network, and it is a crucial problem to reduce the size of giant connected component fast, even to break it down, thus this problem is regarded as finding the minimal set of nodes to reduce the giant component as fast as possible \cite{albert2000error, cohen2001breakdown}.
\begin{figure}[htbp]
\begin{center}
\includegraphics[scale=0.3]{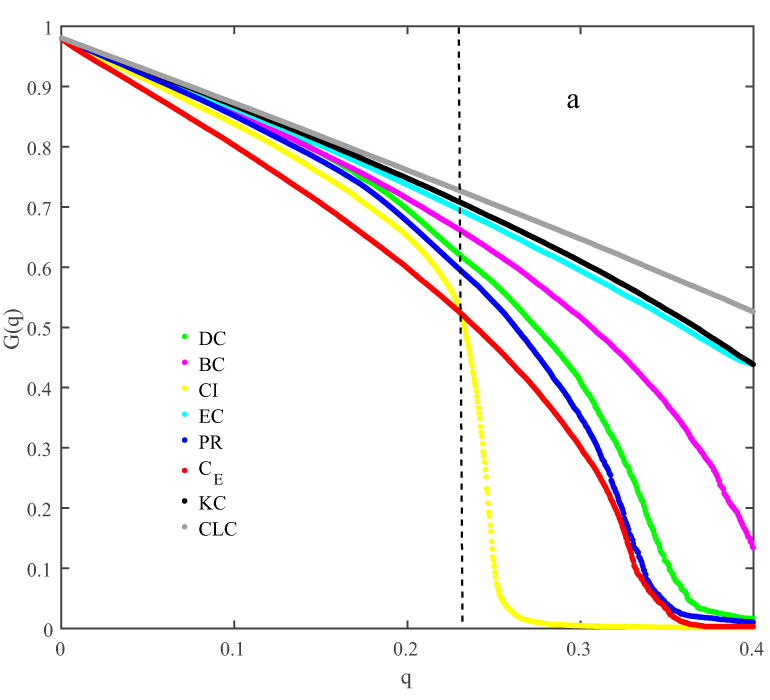}
\includegraphics[scale=0.3]{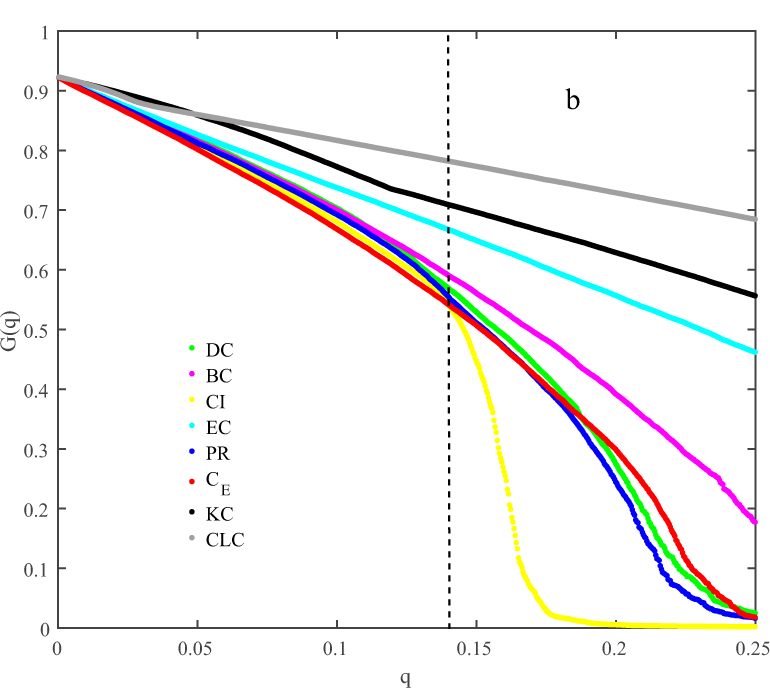}
\caption{\label{Fig:4} \textbf{(a.)} $G(q)$ in ER network. The results are the average values of 20 ER networks. Each network contains 5,000 nodes and 10,000 edges. The performances of degree centrality, betweenness centrality, collective influence, eigenvector centrality, PageRank, k-core, clustering coefficient and von Neumann entropy centrality are represented in different colors. \textbf{(b.)} $G(q)$ in SF network. The points are the average of 20 network with 5,000 nodes and $\gamma = 2.5$.}
\end{center}
\end{figure}

Considering a network with $N$ nodes, when a fraction $q$ of total nodes are removed, the size of the giant component is denoted as $G(q)$. When applying our entropy centrality, at first, node $v_i$ with highest centrality $C_E(v_i)$ is removed and the degrees of neighbors of node $v_i$ are decreased by one. Then in the new network, the same process is implemented again to find a new node with the highest $C_E$ to remove.

Figure \ref{Fig:4}\textbf{a} shows the variations of $G(q)$ as fraction $q$ nodes are removed in Erd\"os-R\'enyi (ER) networks. The von Neumann entropy centrality is compared to several other centralities on the same networks: degree centrality, betweenness centrality, eigenvector centrality (EC) \cite{freeman1978centrality}, Page Rank (PR), k-core (KC) \cite{kitsak2010identification}, clustering coefficient (CLC) \cite{luce1949method} and collective influence (CI) \cite{morone2015influence}. It is illustrated that before $q$ reaches $0.2$, von Neumann entropy centrality outperforms all the other ones. When $q$ reaches $0.2$, $G(q)$ of CI performs a rapid decrease and overpasses $C_E$. Similar results are shown in scale-free (SF) \cite{newman2010networks} networks in Figure \ref{Fig:4}\textbf{b}.

In the viewpoint of centrality, the aim is to identify the most influential nodes on huge networks. To rapidly reduce the scale of giant connected component, our entropy centrality has a wonderful performance. Actually, CI is such an excellent algorithm in solving the optimal influencer problem and it is really hard to be beaten, yet we could observe that the $C_E$ has its own advantages at the beginning when the $q$ is low. Breaking down the giant connected component thoroughly is a long-term goal. Compared to this, reducing its size as fast as possible is a much more practical and realistic target. According to the performances, the entropy centrality is the fastest localized greedy strategy in reducing the size of giant component in some parameter ranges.

\subsection{Average Clustering Coefficient}
Another fascinating phenomenon related to von Neumann entropy centrality is the variation in average clustering coefficient. The global average clustering coefficient of a network is defined as $\bar{C} = \frac{1}{n}\sum_{i=1}^nC_i$, where $C_i$ is the clustering coefficient of node $v_i$
\begin{eqnarray}
C_i = \frac{2|\{(v_j,v_k)\in E:(v_i,v_j)\in E, (v_i,v_k)\in E\}|}{d_i(d_i-1)},
\end{eqnarray}
which indicates how well the neighbors of node $v_i$ are connected.

\begin{figure}[htbp]
\includegraphics[scale=0.3]{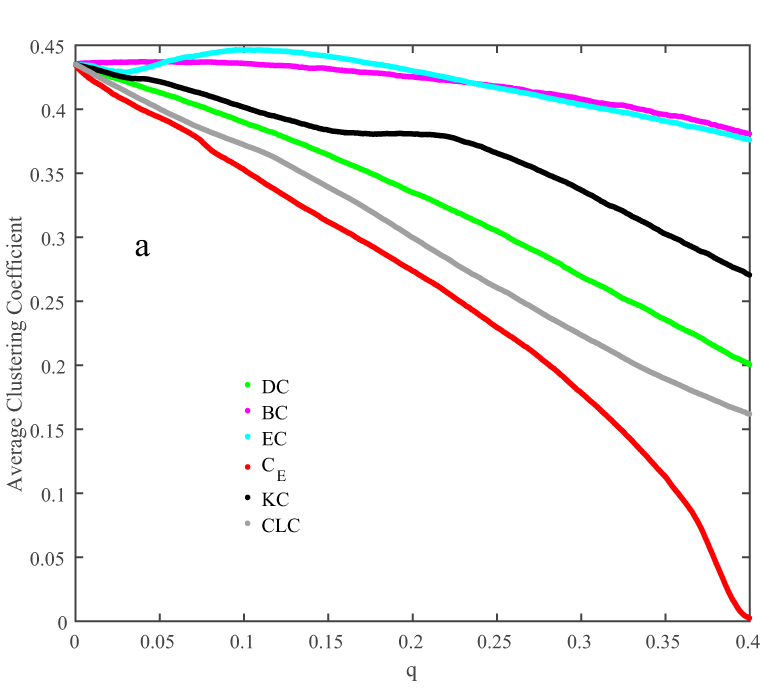}
\includegraphics[scale=0.16]{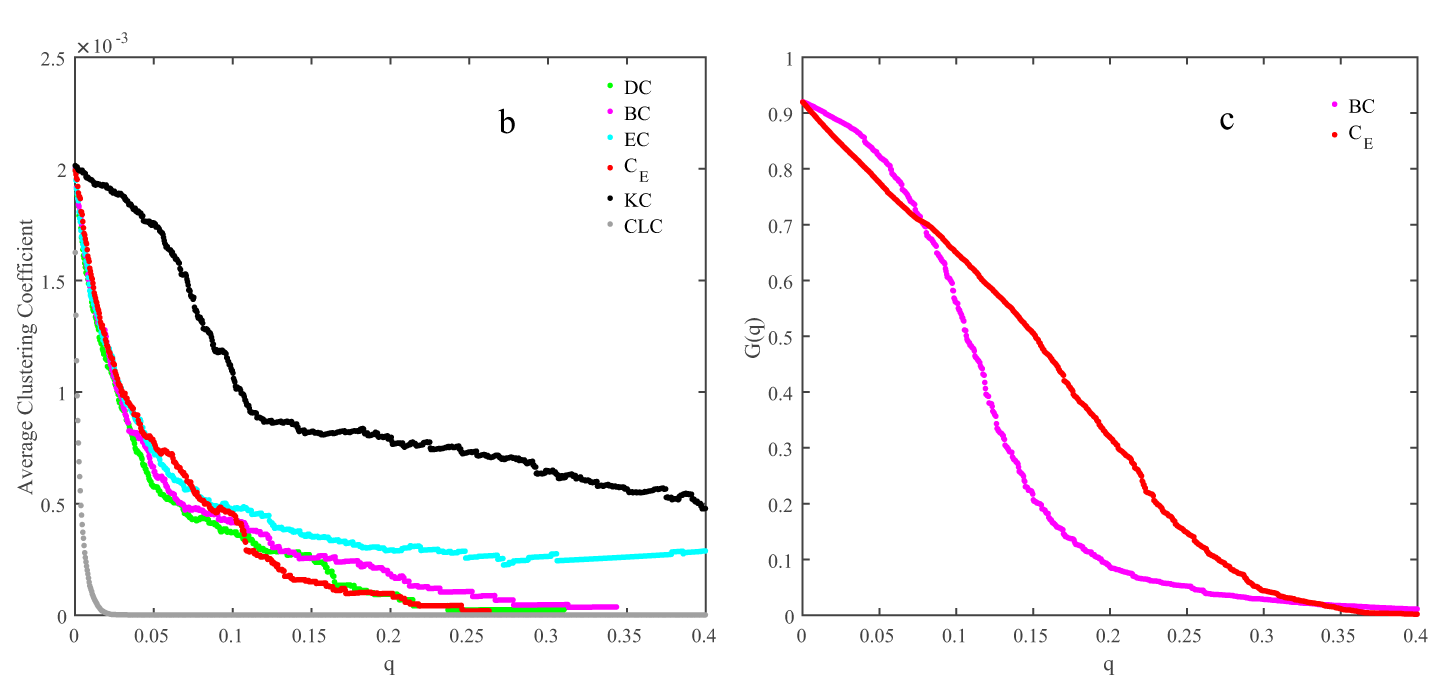}
\begin{center}
\caption{\label{Fig:5} \textbf{(a.)} Average clustering coefficient in RGGs. The results are the average values of 20 RGGs. Each network contains 5,000 nodes scattered in a $3$-dimension space and the average degree is 4. The performances of degree centrality, betweenness centrality, eigenvector centrality, k-core, clustering coefficient and von Neumann entropy centrality are represented in different colors. \textbf{(b.)} Average clustering coefficient in SF network. The points are the average of 20 network with 5,000 nodes and $\gamma = 2.5$. \textbf{(c.)} $G(q)$ in RGGs. The results of betweenness centrality and entropy centrality are presented.}
\end{center}
\end{figure}

In the random geometric graphs (RGGs) \cite{penrose2003random}, every time a nodes with the highest $C_E$ is removed, the global average clustering coefficient is calculated. We find that in the comparison with other centralites including PR, DC, CC, CLC, EC, the von Neumann entropy centrality brings a much more rapid decrease of $\bar{C}$ (Figure \ref{Fig:5}\textbf{a}). It is worth noting that the reduction caused by $C_E$ is even more significant than by CLC itself, which suggests that the removal by von Neumann entropy centrality brings more structural damages than others to RGGs. As a comparison, this phenomenon does not appear in SF network (Figure \ref{Fig:5}\textbf{b}).

Actually, this phenomenon is deeply related to the special topological features of RGGs. The RGGs are the networks whose vertices are scattered randomly in $d$-dimension space. If the distance between two nodes is less than a specific threshold $r$, then these two nodes are linked. One of the most important property of RGGs is that the cluster or modularity structure is striking and there are a lot of large or small clusters in each RGG. Nodes inside each cluster are densely connected and less connected to outliers. This point is also supported by the significance of BC in reducing the size of giant component $G(q)$ (Figure \ref{Fig:5}\textbf{c}) since the BC breaks down the giant components fast, which means there are a few nodes working as bridges between clusters and own highest BC values.

\begin{figure}[htbp]
\begin{center}
\includegraphics[scale=0.5]{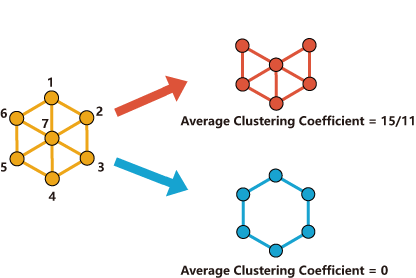}
\caption{\label{Fig:6} An example of cluster. Node $v_7$ owns the lowest clustering coefficient, yet removing node $v_7$ will decrease the $\bar{C}$ to 0.}
\end{center}
\end{figure}

Figure \ref{Fig:6} shows a cluster composed of seven nodes. By definition, clustering coefficients of nodes $v_1$ to $v_6$ are all $\frac{2}{C^2_3}=\frac{2}{3}$ and node $v_7$ is $\frac{6}{C_6^2}=\frac{2}{5}$. Yet when node $v_7$ which owns the highest $C_E$ is removed, the $\bar{C}$ of this whole cluster is brought down to zero. That's the reason why the entropy centrality causes larger reduction in average clustering coefficient than CC to RGGs in the experiments. Also, this phenomenon suggests that $C_E$ does obtain the centre nodes or hubs in the networks efficiently and is able to break down the cluster structures rapidly.

\subsection{Spearman's Rank Correlation Coefficient}
\begin{figure}[htbp]
\includegraphics[scale=0.3]{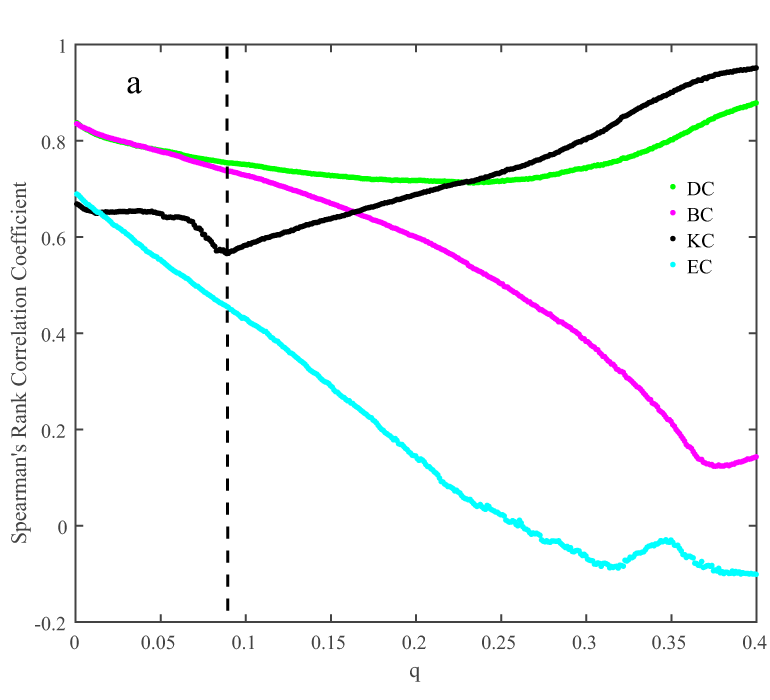}
\includegraphics[scale=0.16]{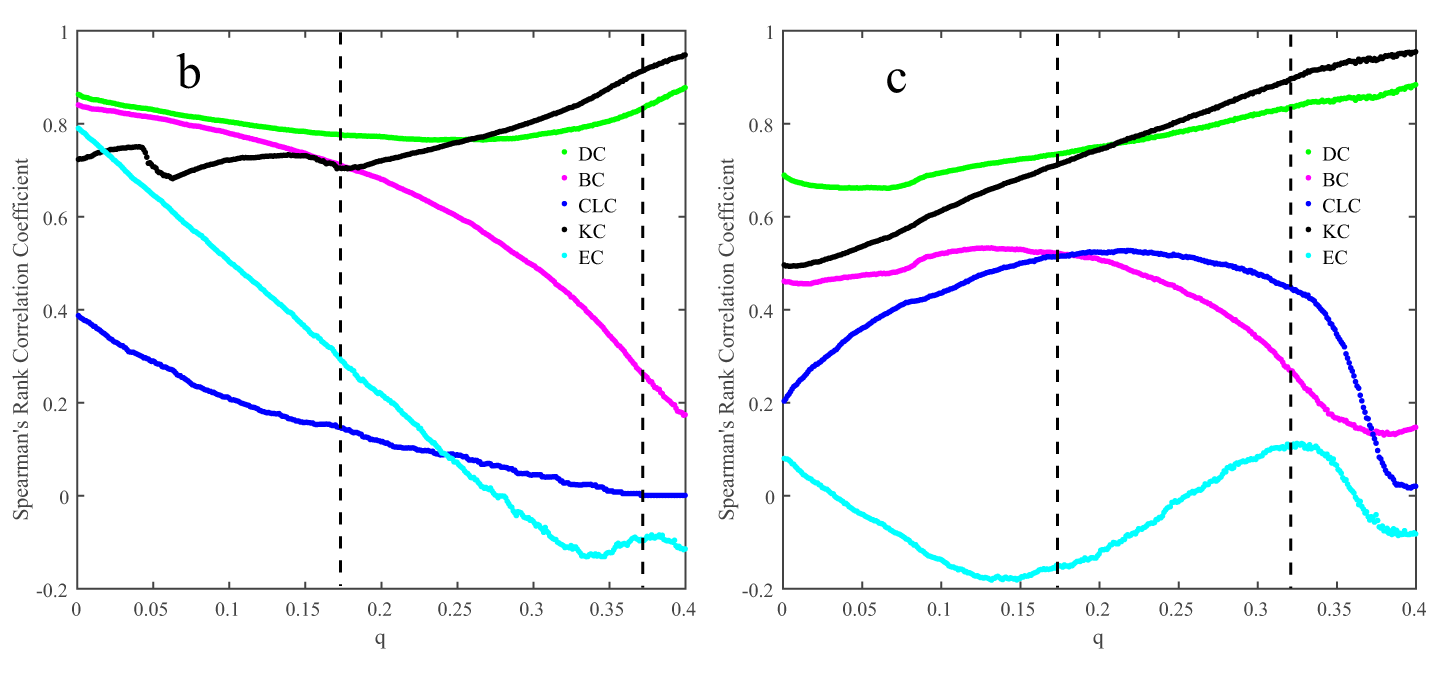}
\begin{center}
\caption{\label{Fig:7} \textbf{(a.)} Spearman's rank correlation coefficient in ER. The results are the average values of 20 ERs. Each network contains 1,000 nodes and the average degree is 4. The correlation coefficient between entropy centrality and degree centrality, betweenness centrality, eigenvector centrality, k-core are represented in different colors. The line is where KC starts to get higher. \textbf{(b.)} Spearman's rank correlation coefficient in SF network between entropy centrality and DC, BC, EC, KC, CLC. The results are the average of 20 network with 1,000 nodes and $\gamma = 2.5$. The left line is where KC starts to get higher. The right line is where CLC touches zero, which mean there is no triangle structure remained. \textbf{(c.)} Spearman's rank correlation coefficient in RGGs. There are 1,000 nodes which are scattered in a 3-dimension space in each network. The BC is more significant in RGGs than in other networks. The left line is where BC starts to get higher and CLC gets steady. The right line is where CLC starts to go down, which means the triangle structures are disappearing.}
\end{center}
\end{figure}

With nodes deleted in entropy centrality order, the variations in the patterns of remained nodes sequences are worth a review. Here the Spearman's rank correlation coefficient is used to further explain and review the entropy centrality. The Spearman correlation coefficient is a measure of rank correlation, which evaluates how well the correlation between two variables could be represented by a monotonic function \cite{myers2013research}. For two sequences $X$ and $Y$, let the ranks sequences of their entries to be $R(X)$ and $R(Y)$, the Spearman correlation coefficient $r_s$ is calculated as:
\begin{eqnarray}
r_s(X,Y) = \frac{Cov(R(X),R(Y))}{\sigma_{R(X)}\sigma_{R(Y)}}.
\end{eqnarray}
In the experiments, every time a node is removed, the Spearman correlations between entropy centrality and DC, BC, CLC, EC and KC are calculated. The results are shown in Figure \ref{Fig:7}.

Figure \ref{Fig:7}\textbf{a} shows the changes of Spearman correlation coefficients between entropy centrality and others in ER networks. As we could see, at the very beginning, major centralities like DC, BC and EC, perform high correlations with $C_E$. This suggests that the entropy centrality could capture the crucial features related to complex networks and $C_E$ could find hubs effectively in the whole networks. With nodes being deleted, the correlation between $C_E$ and $BC$ goes down, which shows the focus of entropy centrality is shifting from finding the bridge nodes between clusters to finding large degrees hubs. This also suggests that as nodes with high entropy centrality are deleted, the large components and clusters get decomposed and the bridge functions of nodes get less significant in the perspective of entropy centrality. The breaking down of giant component could also be illustrated by the increasing of correlation between entropy centrality and KC: the complicated clustering structures disappear and all nodes own low k-core numbers. The whole network degenerates into small pieces with little triangle structures. Since its connection is rare, the centralities requiring a proper global connectivity, like BC and EC, get less significant.

The Spearman correlation coefficients in SF (Figure \ref{Fig:7}\textbf{b}) also perform the same properties. Comparing to ER, at the beginning all the coefficients are higher. This could be explained by features of SF: since there exist high-degree nodes, the crucial hubs in the perspective of many centralities are quite similar. As the number of remained nodes decrease, the CLC goes down to zero, which means the triangle structures disappear and the network exhibits a tree-like shape. The relatively high correlation between $C_E$ and BC in RGGs (Figure \ref{Fig:7}\textbf{c}) shows the high significance of BC in the geometric networks, which coincides with the results of average clustering coefficients and $G(q)$ in RGGs.

\section{Conclusion and Discussion}
In this paper the node centrality based on von Neumann entropy is discussed, which makes it possible to study the importance of nodes in the perspective of structural complexity and features. By comparing the entropy node centrality with classical node centrality, it is shown that the $C_{E}$ is an all-round measurement of node importance. By comparing the changes of size of giant component and average clustering coefficient with other centrality indices when deleting high $C_{E}$ nodes, it is concluded that the von Neumann entropy centrality has an excellent performance in breaking down network structure and can capture the significant features.

Another advantage of von Neumann entropy centrality is that this definition could be expanded to mesoscopic subjects, like motifs. In 2002, Alon et al.\ \cite{shen2002network} introduced the idea of motif when they were studying the gene network, which is defined as the recurring, significant sub-networks and patterns in a network, and it is discovered that the frequencies of some specific motifs in realistic networks are much more significant by comparing with random networks \cite{milo2002network}. Since motifs emphasize on the structure and connection patterns which could not be found by only observing single nodes, node centralities could not capture the structural characterizations completely. Also, for many node centralities, like eigenvector centrality and closeness centrality, they are hard to be generalized to motifs directly. The von Neumann entropy provides an access to evaluate and measure the impact of specific structure on the global network and a new perspective to study network structural features.

Since a great number of real-world data is directed, it is worth defining and researching the von Neumann entropy on directed networks. Chung provided a definition of Laplacian matrix on directed networks \cite{chung2005laplacians} using Perron-Frobenius Theorem \cite{Horn:2012:MA:2422911} and based on this work, Ye et al.\ \cite{ye2014approximate} proposed a method to approximate the von Neumann entropy of directed networks, which allows us to compute the von Neumann entropy in terms of in-degree and out-degree of nodes simply. However, these results only work on strongly-connected directed networks. Another definition involving incidence matrix \cite{godsil2013algebraic}, loses the direction information when calculating the Laplacian. It is still an open problem to define the von Neumann entropy on directed networks generally.

\section*{Acknowledgements}
This work is supported by the Fundamental Research Funds for the Central Universities, the National Natural Science Foundation of China (No.11201019), the International Cooperation Project No.2010DFR00700 and Fundamental Research of Civil Aircraft No.MJ-F-2012-04.

\end{document}